\documentclass[12pt,preprint]{aastex}

\usepackage{graphicx}
\usepackage{wasysym}
\usepackage{apjfonts}
\usepackage{natbib}
\usepackage[colorlinks,urlcolor=cyan,citecolor=blue,linkcolor=blue]{hyperref}

\shortauthors{Terrestrial Planets Orbiting Ultra-Cool Stars and Brown Dwarfs, B.-O. Demory et al.} 
\shorttitle{Terrestrial Planets Orbiting Ultra-Cool Stars and Brown Dwarfs, B.-O. Demory et al.}
\begin{document}

\title{Searching for Terrestrial Planets Orbiting in the Habitable Zone \\ of Ultra-Cool Stars and Brown Dwarfs $^{\star}$}

\author{
Brice-Olivier~Demory\altaffilmark{1}, 
Sara~Seager\altaffilmark{1}, 
Jack~Lissauer\altaffilmark{2}, 
Gregory~Laughlin\altaffilmark{3},
Daniel~Huber\altaffilmark{2},
Matthew~Payne\altaffilmark{4},
Amaury~Triaud\altaffilmark{5},
Michael~Gillon\altaffilmark{6},
Julien~de~Wit\altaffilmark{1},
Andras~Zsom\altaffilmark{1},
Vlada~Stamenkovic\altaffilmark{1},
Franck~Selsis\altaffilmark{7},
J\'er\'emy~Leconte\altaffilmark{8} and
Didier~Queloz\altaffilmark{9}
}

\altaffiltext{1}{Department of Earth, Atmospheric and Planetary Sciences, Massachusetts Institute of Technology, 77 Massachusetts Ave., Cambridge, MA 02139, USA. demory@mit.edu}
\altaffiltext{2}{NASA Ames Research Center, Moffett Field, CA 94035, USA.}
\altaffiltext{3}{UCO/Lick Observatory, Department of Astronomy \& Astrophysics, University of California at Santa Cruz, Santa Cruz, CA 95064, USA}
\altaffiltext{4}{Harvard-Smithsonian Center for Astrophysics, 60 Garden St., Cambridge, MA 02138, USA}
\altaffiltext{5}{Kavli Institute for Astrophysics \& Space Research, Massachusetts Institute of Technology, Cambridge, MA 02139, USA}
\altaffiltext{6}{Institut d'Astrophysique et de G\'eophysique, Universit\'e de Li\`ege, All\'ee du 6 Ao\^ut, 17, Bat. B5C, Li\`ege 1, Belgium.}
\altaffiltext{7}{Universit\'e de Bordeaux, Observatoire Aquitain des Sciences de l'Univers, BP 89, 33271 Floirac Cedex, France}
\altaffiltext{8}{Laboratoire de M\'et\'eorologie Dynamique, Institut Pierre Simon Laplace, CNRS, 4 place Jussieu, BP99, 75252, Paris, France}
\altaffiltext{9}{Astrophysics Group, Cavendish Laboratory, JJ Thomson Avenue, Cambridge CB3 0HE, United Kingdom}
\altaffiltext{$^{\star}$}{White paper submitted in response to the {\it Kepler} Science Office's Call for White Papers (CWP) published at \url{http://keplergo.arc.nasa.gov/docs/Kepler-2wheels-call-1.pdf}}

\begin{abstract}

We propose to use {\it Kepler} in 2-wheel mode to conduct a detailed search for Earth-sized planets orbiting ultra-cool stars and brown dwarfs (spectral types from M7 to L3). This population of objects presents several advantages for exoplanet surveys. First, ultra-cool stars and brown dwarfs are small and thus result in favorable planet-to-star area ratios. Second, because of their low effective temperature, the inner edge of their habitable zone is extremely close (2 to 3 days only). Third, our targets are bright at infrared wavelengths, which will enable detailed follow-up studies. Our program therefore represents a unique opportunity to find a transiting Earth-size exoplanet for which atmospheric features (including biosignatures) could be detected with near-to-come facilities such as {\it JWST}. Such exoplanet has not been discovered yet. {\it Kepler} in 2-wheel mode is the only facility that provides the required stability and photometric precision to make this survey successful. Our initial target sample includes 60 ultra-cool stars and brown dwarfs from which we expect to detect at least one transiting planet. We propose to monitor each source for 4 days, resulting in a total program duration of $\sim$240 days. 

\end{abstract}

\section{Proposed Science}

\subsection{Transiting Planets \--- From Sun-like Stars to Brown Dwarfs}

During its 3.5-year prime mission, the {\it Kepler} mission has revolutionized the field of exoplanets. With 134 confirmed planets and more than 5,000 planet candidates to date, the {\it Kepler} harvest has revealed an astounding and unexpected diversity of planetary properties and environments. One of the main highlights from the mission is the finding that about 50\% of sun-like stars harbor super-Earth planets with orbital periods less than 85 days \citep{Howard:2012,Fressin:2013}. This important result is found to be consistent with radial-velocity surveys \citep{Mayor:2009,Howard:2010,Mayor:2011b,Wolfgang:2011a} employing different instruments and target samples. While {\it Kepler} was primarily designed to optimize the search for planets orbiting sun-like stars, about a thousand of cool, M-dwarf stars have also been monitored. This valuable sample allowed \citet{Dressing:2013} to determine that the occurrence rate of planets around M dwarfs is almost 1, meaning that nearly all M dwarfs harbor at least one planet. This is extremely promising, considering that M dwarfs represent 75\% of the total stellar population in our galaxy. {\bf Billions of planets orbiting cool stars are therefore waiting to be discovered}. 

In the search for terrestrial transiting planets, cool stars are extremely promising because of their small sizes of 0.5 to 0.1 $R_{\odot}$ \citep{Nutzman:2008,Berta:2013}. An Earth-size planet orbiting a late, ultra-cool M dwarf (M8) yields a transit depth 100 $\times$ deeper than an Earth-size planet transiting the Sun. Terrestrial planets orbiting cool stars are therefore easier to characterize, as demonstrated by \citet{Bean:2010} who obtained a ground-based transmission spectrum of GJ\,1214\,b , a 2.7 $R_{\oplus}$ super-Earth orbiting a 0.16 $M_{\oplus}$ M4.5 star \citep{Charbonneau:2009}. Such observations for a similar planetary size would not have been possible for a solar host. Ultra-cool stars (defined here as spectral type cooler than M6) are therefore extremely attractive targets for the search and the characterization of terrestrial planets, with sizes similar to our Earth. This is now widely acknowledged in the exoplanet community as shown by the growth of projects focusing on this population of stars.

Extrapolating this logic, we consider planet formation around brown dwarfs. While ultra-cool stars are just above the hydrogen burning minimum mass, brown dwarfs start just below it and thus are also natural targets to search for planets. A search for planets orbiting brown dwarfs is supported by several instances of circumstellar disks that have been discovered around young brown dwarfs \citep[e.g.,][]{Luhman:2005,Scholz:2008}. While brown dwarfs are expected to accrete only a small amount of material, the total mass could be sufficient to enable planet formation \citep{Payne:2007} and possibly form one to two super-Earths planets. Another similarity with stars is that grain growth in disks around brown dwarfs has been recently shown to be akin to T Tauri stars \citep{Meru:2013,Pinilla:2013}. Very recently a microlensing planet orbiting a brown dwarf has been announced \citep{Han:2013}. This is an excellent motivation to extensively search for planets within the population of close brown dwarfs. 

{\bf In this white paper, we propose to monitor M7 to L3 type objects with {\it Kepler} to search for planets having similar sizes to the Earth and orbiting in the habitable zone of their host star. Our target sample made of nearby, bright sources will enable detailed follow-up observations, including the prospects of detecting atmospheric features with JWST.} We will be searching for potentially habitable planets amongst our closest neighbors.

Interestingly the architecture of planetary systems around ultra-cool stars and brown dwarfs may share similarities. Indeed, it has been theorized that planet formation around ultra-cool stars is expected to be as frequent as around solar-type stars, albeit with smaller planet-star separations \citep{Pascucci:2011}. From a theoretical standpoint, such configurations favor nearly coplanar and closely packed systems of terrestrial planets \citep[e.g.,][]{Montgomery:2009a} that have substantial transit probabilities. Such a scenario is remarkably illustrated by the two exoplanet systems orbiting cool stars found to date, GJ\,1214 \citep{Charbonneau:2009} and Kepler-42 \citep{Muirhead:2012}, for which all planets are orbiting their host star in less than 2 days. \citet{Chiang:2013} recently discussed the similarities between the population of super-Earths orbiting low-mass main sequence stars and moons orbiting our Solar System's gas and ice giants. The similar architecture of these two populations could reveal similarities in the formation mechanism. All of our giant planets have moons and 50\% of our neighborhood solar-like stars harbor super-Earths. We could therefore interpolate this reasoning into the ultra-cool star and brown-dwarf mass regime, which would also host a vast population of exoplanets with periods as short as a few hours.

While there is only a marginal change in radius from the coolest M dwarfs to the hottest brown dwarfs (early L spectral type), the main difference between these two categories of objects is the significant temperature drop. One consequence is that the habitable zone around brown dwarfs is even closer than for ultra-cool stars and can be as short as two days only \citep{Bolmont:2011}, thus dramatically increasing the transit probability of potentially habitable planets.

\subsection{From the Habitable Zone to Habitable Planets}

It is safe to claim that the holy grail of exoplanet research is to find an Earth-like planet that is habitable and that shows signs of life. The habitable zone is the orbital range around a star within which surface liquid water could be sustained. Since water is essential for life as we know it, the search for biosignature gases naturally focuses on planets located in the habitable zone of their host stars. The variety of recent studies on habitability \citep[see, e.g.,][for a review]{Seager:2013} has revealed that low extreme ultra-violet (UV) radiation environments increase the lifetime and concentration of biosignature gases in planetary atmospheres. On the other hand, extreme UV radiation generates reactive radical species such as HO$^{-}$ that breaks chemical bonds of several atmospheric gases. Because of their cooler temperature, quiet M stars radiate significantly less flux at extreme UV than for a typical G dwarf like our Sun. Rocky planets orbiting ultra-cool stars therefore represent the best prospects for finding biosignature gases in their atmospheres. Exoplanets orbiting ultra-cool stars have an additional advantage besides their promise for habitability. Because of their lower effective temperature, ultra-cool stars harbor an habitable zone that is much closer to the host star than that of a solar-type star, which increases both the probability of transit as well as the number of transits that can be observed in a given timeframe. {\bf Terrestrial planets orbiting in the habitable zone of ultra-cool stars do not only yield larger transit depths than for sun analogues but are also confirmed more rapidly.}

The inner edge of the habitable zone is potentially much closer to the host star than previously thought. \citet{Zsom:2013} recently found that planets with a limited water reservoir are habitable closer to the host star than an Earth-analog planet. For example, a planet with 1 bar N$_2$ atmosphere, 1\% relative humidity and a surface albedo of 0.2 is habitable as close as 0.015 AU from an M6 host. If the surface albedo is larger, 0.8, the minimum habitable distance is expected to be as low as 0.01 AU. In comparison, an Earth analog habitable planet has to be located further out than 0.03 AU around the same host \citep{Kopparapu:2013}.

\noindent {\bf How frequent are planets orbiting in the habitable zone of ultra-cool stars and brown dwarfs?} Based on the results of \citet{Zsom:2013}, the occurrence rate of terrestrial planets (with radius between 0.5 and 2 R$_{\oplus}$) in the Habitable Zone for M dwarfs is around 0.7-0.9. In comparison, it is 0.5-0.6 based on the Earth-analog Habitable Zone limits of \citet{Kopparapu:2013}. The occurrence rate is $0.70^{+0.15}_{-0.09}$ assuming a surface albedo of 0.2, and $0.88^{+0.09}_{-0.05}$ assuming a high surface albedo case of 0.8. \citet{Zsom:2013} applied the method of \citet{Dressing:2013} to estimate the occurrence rate and assume that the outer edge of the Habitable Zone is at infinity. For an M6 host, the inner edge of the habitable zone is located below 2 days (for 0.8 albedo) and 2.6 days for the low albedo case (Figure~\ref{figwp} (left)). As we plan to monitor each target during $\sim$4 days (see below), our observations focusing on the cooler M7 to L3 hosts will comfortably encompass their inner habitable zone.

\noindent {\bf Will JWST study an habitable world?} Several studies have demonstrated that the James Webb Space Telescope \citep[JWST,][]{Gardner:2006} will have the opportunity to detect atmospheric features in the transmission spectrum of a rocky exoplanet only if it orbits an M dwarf \citep{Belu:2011}. And even in this case, this would require observing all eclipses during JWST mission. {\bf Cooler hosts, such as the sample we propose to target in this program, are the only ones for which detecting atmospheric features (including biosignatures) in small exoplanets will be achievable in a reasonable amount of JWST time.} A simulated spectrum for an Earth-like exoplanet orbiting a brown dwarf is shown on Figure~\ref{figwp} (right).

\begin{figure}[!ht]
\epsscale{1.15}
\plottwo{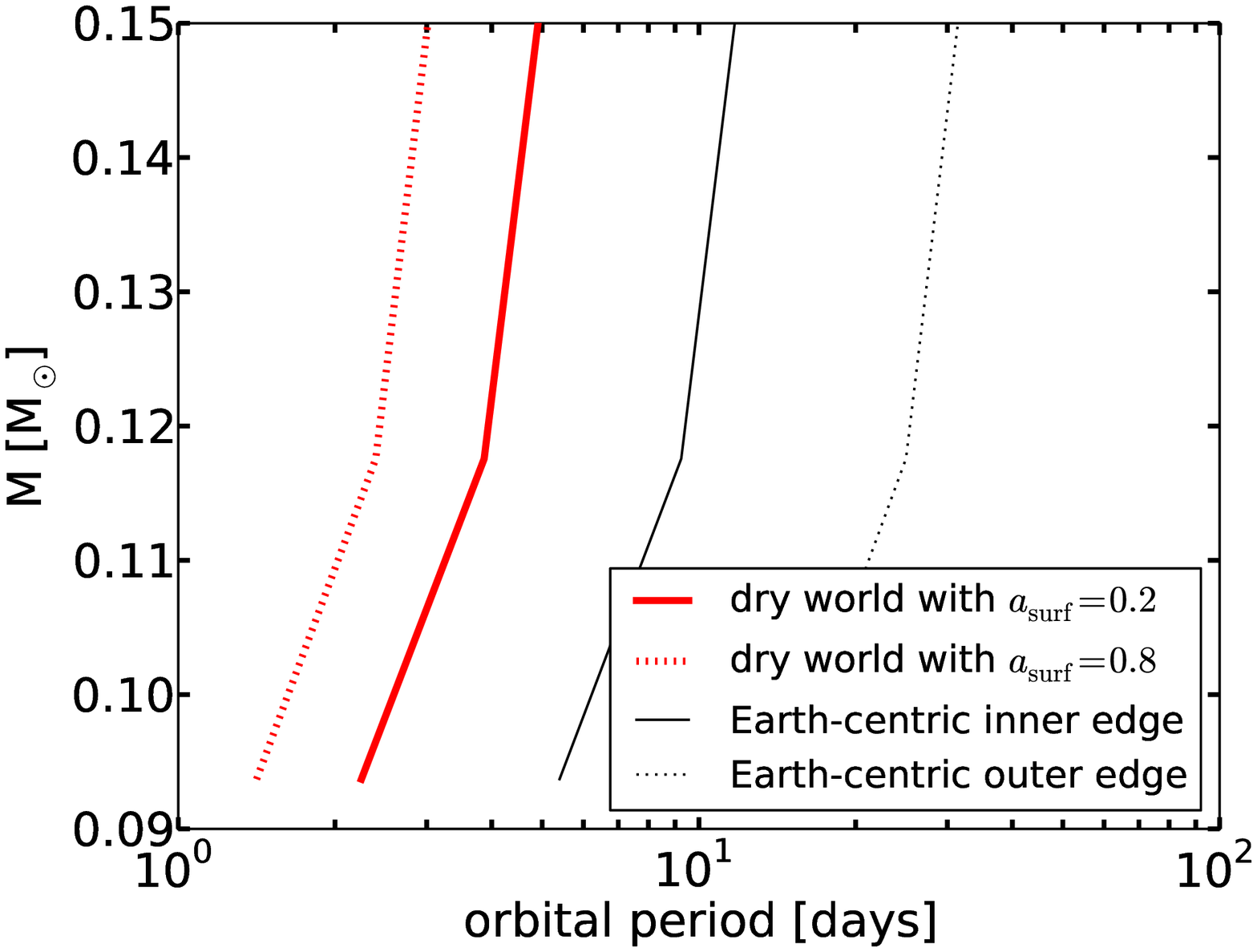}{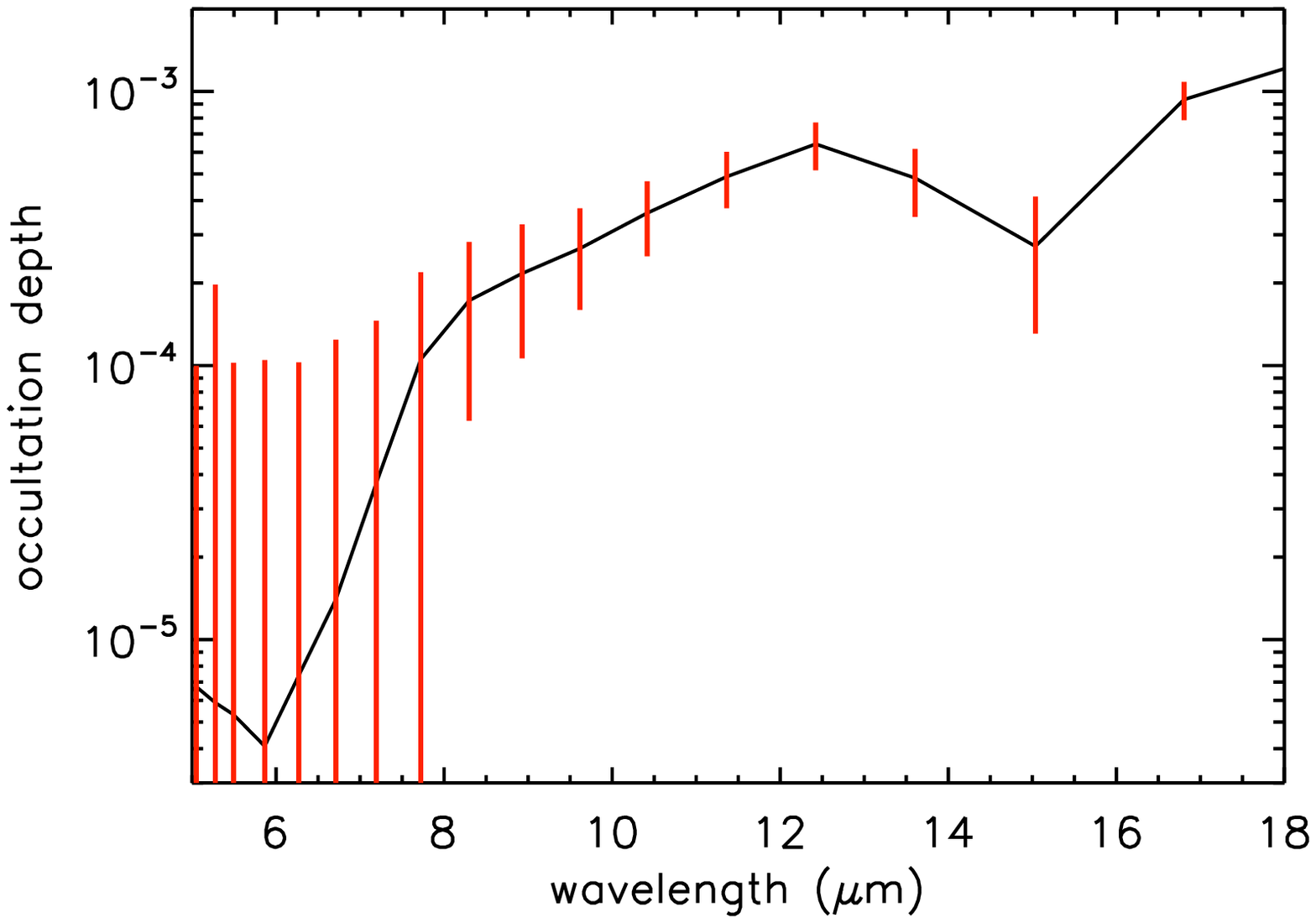}
\caption{{\bf Left:} Orbital periods in the habitable zone of low mass stars. The x axis is the orbital period in days, the y axis is the stellar mass in units of solar mass. The red dotted line shows the orbital period for a habitable planet with 1 bar N$_2$ atmosphere, high surface albedo (0.8) and low relative humidity (1\% in the troposphere). The red solid line corresponds to a more plausible lower surface albedo scenario (0.2). The orbital periods at the inner and outer edges of the Earth-centric habitable zone \citep{Kopparapu:2013} is also shown for comparison by the black solid and dotted lines, respectively.  {\bf Right:} Simulated JWST emission spectrum of a planet similar to Earth but with an atmosphere devoid of O$_3$. The thermal profile is determined by irradiation from a 60 M$_{{\rm Jup}}$ brown dwarf host (Teff=1600 K at 1.5 Gyr) located at 10pc. Here, 50 occultations are stacked together and the error bars are 5-$\sigma$ uncertainties. One of the CO$_2$ absorption bands is confidently detected. Note that for M dwarfs or brown dwarfs emitting UV radiation (which is relevant for our program), a clear O$_3$ absorption band would also be visible at 9.6 $\mu$m, provided the planet atmosphere is O$_2$ rich. } \label{figwp}
\end{figure}

\subsection{Science Goals \--- Finding A Rocky Exoplanet Amenable to Characterization}

Our primary science goal is to find terrestrial planets orbiting bright nearby, ultra-cool stars and brown dwarfs. Our list encompasses $\sim$60 bright targets with stellar types from M7 to L3. This spectral range has been mostly left unexplored by transit surveys. All targets benefit from spectroscopic observations that will facilitate the characterization of planets transiting them. None of the small planets found so far with {\it Kepler} will be amenable to detailed characterization, even with JWST. {\bf Our program represents a unique opportunity to find a rocky planet for which a transmission/emission spectrum could be obtained with near-to-come facilities such as JWST.}

Finding Earth-size planets that orbit ultra-cool stars is possible using {\it Kepler} in its 2-reaction wheel mode and assuming the photometric performance described in the Call for White Papers (see Technical section below). Thanks to 1) ultra-cool stars and brown dwarfs' small sizes (hence favorable planet-to-star area ratio), 2) the proximity of their habitable zone and 3) their brightness at infrared wavelengths, these objects represent our best opportunity to find terrestrial planets in which biosignatures could be detected with the next generation facilities such as JWST.

{\bf How many planets will be found?} To estimate the expected yield from our observations, we consider the {\it Kepler} Objects of Interest (KOI) based on Q0-Q12. In the Kepler Input Catalog (KIC) there are approximately $\sim$100 dwarf\footnote{We cut targets with J-K > 1.0, which eliminates objects that are likely background giant stars.} stars with Kepmag $<17.5$ and $r-K > 4.5$, corresponding to an effective temperature threshold of an M6 star (3,100\,K). Cross-matching these stars with the most recent list of KOIs yields two planetary systems: Kepler-42 and KOI-2704. We can therefore expect to find one planet in our sample of 60 targets. The yield will be even higher if we consider that small planets are likely to be found in multi-planetary systems. We note that the {\it Kepler} pipeline does not search for planetary candidates with periods less than 0.5 days \citep{Sanchis-Ojeda:2013}, so we consider this expected yield to be a lower threshold.

\subsection{Science Goals \--- Ultra-Cool Star and Brown Dwarfs Physics}

Our proposed high-precision monitoring of ultra-cool stars and brown dwarfs will prove to be particularly useful for understanding the variability nature of these objects, even if no transit is detected throughout the sample.

For the ultra-cool M-star part of the sample, flares and rotational activity will be easily detected, thus allowing us to constrain the stellar magnetic field properties at the bottom of the main sequence. Coupling these observations to archival UV and X-ray monitoring will provide an estimate of the magnetic energy output from these objects and better assess the overall level of activity of ultra-cool stars \citep{Welsh:2007} as well as exploring possible relationship with age \citep{West:2008} and rotational period.

In the brown-dwarf regime, the typical duration of our observations of 2 to 4 days will allow us to monitor several full rotation periods. Indeed, brown dwarfs contract with age and therefore their rotational velocity increase to reach a few hours only. This sample will be particularly valuable when coupled with similar duration {\it Spitzer} observations of brown dwarfs. Coupling infrared and {\it Kepler}, visible photometry will enable us to probe the vertical atmospheric structure of brown dwarfs, similar to the study recently conducted on the nearby T dwarf 2MASS J22282889-431026 \citep{Buenzli:2012} but for L dwarfs. These hotter brown dwarfs have effective temperatures that place them in an exceptionally rich region of condensation phase space \citep{Morley:2012} which could allow us to detect the signature of clouds in visible bandpasses. Assuming the telescope schedule will be public, other teams could arrange simultaneous observations of our targets in other bandpasses, which could be especially relevant for brown dwarfs atmospheric studies. Finally, using an algorithm based on the autocorrelation function \citep[e.g.][]{McQuillan:2013} of brown-dwarf lightcurves may constrain the alignment of their rotation axis with the observer (whether they are viewed edge-on or with the pole towards the observer) thanks to their fast rotation.

\subsection{Target List and Program Duration}

We base our target list of ultra-cool stars and brown dwarfs on a sample of nearby, well characterized objects that have been subject to other studies. This has two advantages: there is no false positives and deriving the planet properties will be easier to perform. All objects are extracted from the \citet{Lepine:2011} all-sky M-dwarf catalog and the {\it Dwarf Archives}. We estimate the {\it Kepler} magnitude from the 2MASS JHK photometry by using the models from \citet{Baraffe:1998}. We keep all sources with Kepmag$<17.5$. We finally included basic pointing constraints from the Call (see below). Our selection yields 60 suitable candidates. 

The period distribution of planets orbiting ultra-cool stars is not known. As of today, we know only two systems with host stars having effective temperature less than 3,200\,K, those are GJ\,1214 \citep{Charbonneau:2009} and Kepler-42 \citep{Muirhead:2012}. The former hosts one super-Earth and the second hosts three sub-Earth sized companions, all orbiting their star in less than 2 days. While this trend may not reflect all planets orbiting small stars, we choose to survey all stars in our sample (which are cooler than both GJ\,1214 and Kepler-42) for twice the time, ie: 4 days, which corresponds to the maximum time during which the pointing can be maintained. The proposed program would therefore have an estimated duration of about 240 days. {\bf This 4-day long monitoring encompasses the inner region of the habitable zone for all of our targets.}

We finally note that it would be perfectly feasible to conduct other programs alongside our proposed observations. This could be easily done by defining beforehand all apertures related to each program for a given field of view.

\subsection{The Program in Context}

Our proposal meets three objectives of the NASA's Origins Program, namely the characterization of exoplanets, study of planetary formation and the search for cosmic life. The proposed effort will be complementary to the Transiting Exoplanet Survey Satellite (TESS) mission by targeting cooler stars that are beyond the reach of TESS. Indeed, our proposed sample (M7 to L3) is made of relatively faint sources and therefore requires a larger aperture than TESS' or any current dedicated ground-based surveys such as MEarth \citep{Nutzman:2008}. In addition, our proposed program is timely as it can provide an exquisite candidate for characterization with JWST and large ground-based telescopes.

\section{Technical Considerations}

The proposed science described in this white paper is well suited to Kepler's 2-wheel operations. We detail in the following an observing strategy to meet our science goals. 

\subsection{Photometric Precision}

An Earth-size planet transiting a M7-L3 star yields a transit depth of the order of 1\% (10,000 ppm). Therefore most adverse effects amplified by the degraded 2-wheel attitude system will not significantly affect the Kepler's performance for this specific program. We will be mainly dominated by the sources' photon noise and the background contribution. We use the photometric precision estimates provided in the Call as a first proxy to estimate the feasibility of our proposed science. The Call mentions that a photometric precision of the order of 1mmag can be reached for a 12th-magnitude star in 1-minute integration. Assuming pure photon noise, this translates to a precision of $\sim$12.5 mmag per minute for a 17.5th-magnitude star, which is the faintest target of our proposed sample. This translates to a photometric precision of 3 mmag per 15min, which is the approximate transit duration of a planet orbiting a $R=0.1 R_{\odot}$ in just 2 days. This precision would yield a 3-$\sigma$ detection for a single transit event.  For the brightest targets of our sample we will be sensitive to planets smaller than Mercury.

\subsection{Improving the Photometric Precision}

The goal of our white paper is to propose science that is readily achievable assuming the preliminary photometric performance detailed in the Call. However, we investigate in the following a few ways to improve the photometry using techniques that have proved to be successful with other facilities. Our team has developed a strong experience about space-based photometry with {\it Kepler}, {\it Spitzer} and {\it HST}.  We propose in the following several ideas to improve photometric precision. 

\noindent {\bf Differential Photometry.} The most straightforward approach to mitigate instrumental systematics that will arise from the degraded attitude control system is to simultaneously monitor stars that are close to the target and to identify the common systematic patterns in their lightcurves. A Bayesian Maximum A posteriori Probability (MAP) approach similar to the one currently employed by the Science Office and described in \citet{Smith:2012a} will provide a set of cotrending basis vectors (CBV) that can be efficiently used to mitigate instrumental systematics such as differential velocity aberration and all temperature driven effects. Temperature effects related to attitude changes are likely to be more important in the 2-wheel operation. The successful implementation of such a Bayesian approach requires a set of quiet and highly-correlated stars which will be used to extract the set of CBV. Because our ultra-cool stars and brown dwarfs targets are faint, background stars of similar Kepler magnitude located in their immediate neighborhood will be numerous, thus providing a large pool of suitable reference stars.

\noindent {\bf Intra-Pixel Sensitivity.} In 2010, we used Q1 HAT-P-7 {\it Kepler} photometry to explore how the centroid position correlates with the low level of systematic noise found in the lightcurves. We expected the effect to be small but not non-existent as Kepler detector's large pixel size undersamples the stellar point response function \citep{Bryson:2010}. Furthermore, a time-varying loss of stellar flux in the pixel-to-pixel interstices could increase the flux-position dependence in the time-series \citep[e.g.,][]{Kinemuchi:2012}. Our results yielded only a nominal correlation, suggesting that intra-pixel sensitivity for {\it Kepler} CCD is negligible during the nominal mission. However, operations in 2-wheel mode will result in significant motion on the detector over short timescales, which might make intra-pixel sensitivity particularly relevant.

\subsection{Pointing constraints}

The {\it Kepler} Science Office recently posted an appendix\footnote{\url{http://keplergo.arc.nasa.gov/docs/Kepler-2-Wheel-pointing-performance.pdf}} to the Call detailing the pointing control in 2-wheel operations. We initially choose to limit our targets' locations within $\pm30$ degrees of the ecliptic. This is an arbitrary constraint that we will able to adjust alongside the limiting magnitude of our sample, pending detailed operational testing. We emphasize that we do not expect position-dependent systematics to be the dominant source of noise for most of our targets. We could therefore accommodate higher rates of drift and more frequent pointing sequences for reaction wheel momentum desaturation.

\subsection{Aperture design and Exposure time}

Apertures will match the path of the target star on the detector during the planned duration of the observations. This will be done by appending apertures each one to another for the target and our reference stars. Assuming a rate of drift of 0.9''/min ($\sim$1 pixel per 5 min, as indicated in the Call), a 4-day continuous monitoring translates to a 1152-pixel-long path on the detector. A 6-pixel wide aperture along this path would result in $\sim$7,000 pixels per target, which could be duplicated for several dozen of reference stars located in the target proximity. 1-min exposure time is adequate for our purposes. Our program does not require modifications to the flight software.

\subsection{Data Reduction and Analysis}

Because of the faintness of our stars, most adverse effects due to the degraded attitude control system will be dominated by photon noise. We therefore expect the data analysis for faint targets will be similar to the one employed with data from the nominal mission.


\begin{thebibliography}{}
\scriptsize

\bibitem[{Baraffe {et~al.}(1998)Baraffe, Chabrier, Allard, \&
  Hauschildt}]{Baraffe:1998}
Baraffe, I., Chabrier, G., Allard, F., \& Hauschildt, P.~H. 1998, \aap, 337,
  403

\bibitem[{Bean {et~al.}(2010)Bean, Kempton, \& Homeier}]{Bean:2010}
Bean, J.~L., Kempton, E. M.-R., \& Homeier, D. 2010, Nature, 468, 669

\bibitem[{{Belu} {et~al.}(2011){Belu}, {Selsis}, {Morales}, {Ribas}, {Cossou},
  \& {Rauer}}]{Belu:2011}
{Belu}, A.~R., {Selsis}, F., {Morales}, J.-C., {et~al.} 2011, \aap, 525, A83

\bibitem[{{Berta} {et~al.}(2013){Berta}, {Irwin}, \&
  {Charbonneau}}]{Berta:2013}
{Berta}, Z.~K., {Irwin}, J., \& {Charbonneau}, D. 2013, ArXiv e-prints

\bibitem[{{Bolmont} {et~al.}(2011){Bolmont}, {Raymond}, \&
  {Leconte}}]{Bolmont:2011}
{Bolmont}, E., {Raymond}, S.~N., \& {Leconte}, J. 2011, \aap, 535, A94

\bibitem[{Bryson {et~al.}(2010)Bryson, Tenenbaum, Jenkins, Chandrasekaran,
  Klaus, Caldwell, Gilliland, Haas, Dotson, Koch, \& Borucki}]{Bryson:2010}
Bryson, S.~T., Tenenbaum, P., Jenkins, J.~M., {et~al.} 2010, \apjl, 713, L97

\bibitem[{{Buenzli} {et~al.}(2012){Buenzli}, {Apai}, {Morley}, {Flateau},
  {Showman}, {Burrows}, {Marley}, {Lewis}, \& {Reid}}]{Buenzli:2012}
{Buenzli}, E., {Apai}, D., {Morley}, C.~V., {et~al.} 2012, \apjl, 760, L31

\bibitem[{Charbonneau {et~al.}(2009)Charbonneau, Berta, Irwin, Burke, Nutzman,
  Buchhave, Lovis, Bonfils, Latham, Udry, Murray-Clay, Holman, Falco, Winn,
  Queloz, Pepe, Mayor, Delfosse, \& Forveille}]{Charbonneau:2009}
Charbonneau, D., Berta, Z.~K., Irwin, J., {et~al.} 2009, Nature, 462, 891

\bibitem[{{Chiang} \& {Laughlin}(2013)}]{Chiang:2013}
{Chiang}, E., \& {Laughlin}, G. 2013, \mnras, 431, 3444

\bibitem[{{Dressing} \& {Charbonneau}(2013)}]{Dressing:2013}
{Dressing}, C.~D., \& {Charbonneau}, D. 2013, \apj, 767, 95

\bibitem[{{Fressin} {et~al.}(2013){Fressin}, {Torres}, {Charbonneau}, {Bryson},
  {Christiansen}, {Dressing}, {Jenkins}, {Walkowicz}, \&
  {Batalha}}]{Fressin:2013}
{Fressin}, F., {Torres}, G., {Charbonneau}, D., {et~al.} 2013, \apj, 766, 81

\bibitem[{Gardner {et~al.}(2006)Gardner, Mather, Clampin, Doyon, Greenhouse,
  Hammel, Hutchings, Jakobsen, Lilly, Long, Lunine, McCaughrean, Mountain,
  Nella, Rieke, Rieke, Rix, Smith, Sonneborn, Stiavelli, Stockman, Windhorst,
  \& Wright}]{Gardner:2006}
Gardner, J.~P., Mather, J.~C., Clampin, M., {et~al.} 2006, Space Science
  Reviews, 123, 485

\bibitem[{{Han} {et~al.}(2013){Han}, {Jung}, {Udalski}, {Sumi}, {Gaudi},
  {Gould}, {Bennett}, {Tsapras}, {Szyma{\'n}ski}, {Kubiak}, {Pietrzy{\'n}ski},
  {Soszy{\'n}ski}, {Skowron}, {Koz{\l}owski}, {Poleski}, {Ulaczyk},
  {Wyrzykowski}, {Pietrukowicz}, {Abe}, {Bond}, {Botzler}, {Chote}, {Freeman},
  {Fukui}, {Furusawa}, {Harris}, {Itow}, {Ling}, {Masuda}, {Matsubara},
  {Muraki}, {Ohnishi}, {Rattenbury}, {Saito}, {Sullivan}, {Sweatman}, {Suzuki},
  {Tristram}, {Wada}, {Yock}, {Batista}, {Christie}, {Choi}, {DePoy}, {Dong},
  {Hwang}, {Kavka}, {Lee}, {Monard}, {Natusch}, {Ngan}, {Park}, {Pogge},
  {Porritt}, {Shin}, {Tan}, {Yee}, {Alsubai}, {Bramich}, {Browne}, {Dominik},
  {Horne}, {Hundertmark}, {Ipatov}, {Kains}, {Liebig}, {Snodgrass}, {Steele},
  \& {Street}}]{Han:2013}
{Han}, C., {Jung}, Y.~K., {Udalski}, A., {et~al.} 2013, ArXiv e-prints

\bibitem[{{Howard} {et~al.}(2010){Howard}, {Marcy}, {Johnson}, {Fischer},
  {Wright}, {Isaacson}, {Valenti}, {Anderson}, {Lin}, \& {Ida}}]{Howard:2010}
{Howard}, A.~W., {Marcy}, G.~W., {Johnson}, J.~A., {et~al.} 2010, Science, 330,
  653

\bibitem[{{Howard} {et~al.}(2012){Howard}, {Marcy}, {Bryson}, {Jenkins},
  {Rowe}, {Batalha}, {Borucki}, {Koch}, {Dunham}, {Gautier}, {Van Cleve},
  {Cochran}, {Latham}, {Lissauer}, {Torres}, {Brown}, {Gilliland}, {Buchhave},
  {Caldwell}, {Christensen-Dalsgaard}, {Ciardi}, {Fressin}, {Haas}, {Howell},
  {Kjeldsen}, {Seager}, {Rogers}, {Sasselov}, {Steffen}, {Basri},
  {Charbonneau}, {Christiansen}, {Clarke}, {Dupree}, {Fabrycky}, {Fischer},
  {Ford}, {Fortney}, {Tarter}, {Girouard}, {Holman}, {Johnson}, {Klaus},
  {Machalek}, {Moorhead}, {Morehead}, {Ragozzine}, {Tenenbaum}, {Twicken},
  {Quinn}, {Isaacson}, {Shporer}, {Lucas}, {Walkowicz}, {Welsh}, {Boss},
  {Devore}, {Gould}, {Smith}, {Morris}, {Prsa}, {Morton}, {Still}, {Thompson},
  {Mullally}, {Endl}, \& {MacQueen}}]{Howard:2012}
{Howard}, A.~W., {Marcy}, G.~W., {Bryson}, S.~T., {et~al.} 2012, \apjs, 201, 15

\bibitem[{{Kinemuchi} {et~al.}(2012){Kinemuchi}, {Barclay}, {Fanelli},
  {Pepper}, {Still}, \& {Howell}}]{Kinemuchi:2012}
{Kinemuchi}, K., {Barclay}, T., {Fanelli}, M., {et~al.} 2012, \pasp, 124, 963

\bibitem[{{Kopparapu} {et~al.}(2013){Kopparapu}, {Ramirez}, {Kasting}, {Eymet},
  {Robinson}, {Mahadevan}, {Terrien}, {Domagal-Goldman}, {Meadows}, \&
  {Deshpande}}]{Kopparapu:2013}
{Kopparapu}, R.~K., {Ramirez}, R., {Kasting}, J.~F., {et~al.} 2013, \apj, 765,
  131

\bibitem[{{L{\'e}pine} \& {Gaidos}(2011)}]{Lepine:2011}
{L{\'e}pine}, S., \& {Gaidos}, E. 2011, \aj, 142, 138

\bibitem[{Luhman {et~al.}(2005)Luhman, Adame, D'Alessio, Calvet, Hartmann,
  Megeath, \& Fazio}]{Luhman:2005}
Luhman, K.~L., Adame, L., D'Alessio, P., {et~al.} 2005, \apj, 635, L93

\bibitem[{Mayor {et~al.}(2009)Mayor, Udry, Lovis, Pepe, Queloz, Benz, Bertaux,
  Bouchy, Mordasini, \& Segransan}]{Mayor:2009}
Mayor, M., Udry, S., Lovis, C., {et~al.} 2009, \aap, 493, 639

\bibitem[{{Mayor} {et~al.}(2011){Mayor}, {Marmier}, {Lovis}, {Udry},
  {S{\'e}gransan}, {Pepe}, {Benz}, {Bertaux}, {Bouchy}, {Dumusque}, {Lo Curto},
  {Mordasini}, {Queloz}, \& {Santos}}]{Mayor:2011b}
{Mayor}, M., {Marmier}, M., {Lovis}, C., {et~al.} 2011, ArXiv e-prints

\bibitem[{{McQuillan} {et~al.}(2013){McQuillan}, {Mazeh}, \&
  {Aigrain}}]{McQuillan:2013}
{McQuillan}, A., {Mazeh}, T., \& {Aigrain}, S. 2013, ArXiv e-prints

\bibitem[{{Meru} {et~al.}(2013){Meru}, {Galvagni}, \& {Olczak}}]{Meru:2013}
{Meru}, F., {Galvagni}, M., \& {Olczak}, C. 2013, ArXiv e-prints

\bibitem[{{Montgomery} \& {Laughlin}(2009)}]{Montgomery:2009a}
{Montgomery}, R., \& {Laughlin}, G. 2009, \icarus, 202, 1

\bibitem[{{Morley} {et~al.}(2012){Morley}, {Fortney}, {Marley}, {Visscher},
  {Saumon}, \& {Leggett}}]{Morley:2012}
{Morley}, C.~V., {Fortney}, J.~J., {Marley}, M.~S., {et~al.} 2012, \apj, 756,
  172

\bibitem[{{Muirhead} {et~al.}(2012){Muirhead}, {Johnson}, {Apps}, {Carter},
  {Morton}, {Fabrycky}, {Pineda}, {Bottom}, {Rojas-Ayala}, {Schlawin},
  {Hamren}, {Covey}, {Crepp}, {Stassun}, {Pepper}, {Hebb}, {Kirby}, {Howard},
  {Isaacson}, {Marcy}, {Levitan}, {Diaz-Santos}, {Armus}, \&
  {Lloyd}}]{Muirhead:2012}
{Muirhead}, P.~S., {Johnson}, J.~A., {Apps}, K., {et~al.} 2012, \apj, 747, 144

\bibitem[{{Nutzman} \& {Charbonneau}(2008)}]{Nutzman:2008}
{Nutzman}, P., \& {Charbonneau}, D. 2008, \pasp, 120, 317

\bibitem[{{Pascucci} {et~al.}(2011){Pascucci}, {Laughlin}, {Gaudi}, {Kennedy},
  {Luhman}, {Mohanty}, {Birkby}, {Ercolano}, {Plavchan}, \&
  {Skemer}}]{Pascucci:2011}
{Pascucci}, I., {Laughlin}, G., {Gaudi}, B.~S., {et~al.} 2011, in Astronomical
  Society of the Pacific Conference Series, Vol. 448, 16th Cambridge Workshop
  on Cool Stars, Stellar Systems, and the Sun, ed. C.~{Johns-Krull}, M.~K.
  {Browning}, \& A.~A. {West}, 469

\bibitem[{{Payne} \& {Lodato}(2007)}]{Payne:2007}
{Payne}, M.~J., \& {Lodato}, G. 2007, \mnras, 381, 1597

\bibitem[{{Pinilla} {et~al.}(2013){Pinilla}, {Birnstiel}, {Benisty}, {Ricci},
  {Natta}, {Dullemond}, {Dominik}, \& {Testi}}]{Pinilla:2013}
{Pinilla}, P., {Birnstiel}, T., {Benisty}, M., {et~al.} 2013, \aap, 554, A95

\bibitem[{{Sanchis-Ojeda} {et~al.}(2013){Sanchis-Ojeda}, {Rappaport}, {Winn},
  {Levine}, {Kotson}, \& {Latham}}]{Sanchis-Ojeda:2013}
{Sanchis-Ojeda}, R., {Rappaport}, S., {Winn}, J.~N., {et~al.} 2013, ArXiv
  e-prints

\bibitem[{{Scholz} {et~al.}(2008){Scholz}, {Jayawardhana}, {Wood},
  {Lafreni{\`e}re}, {Schreyer}, \& {Doyon}}]{Scholz:2008}
{Scholz}, A., {Jayawardhana}, R., {Wood}, K., {et~al.} 2008, \apjl, 681, L29

\bibitem[{{Seager}(2013)}]{Seager:2013}
{Seager}, S. 2013, Science, 340, 577

\bibitem[{{Smith} {et~al.}(2012){Smith}, {Stumpe}, {Van Cleve}, {Jenkins},
  {Barclay}, {Fanelli}, {Girouard}, {Kolodziejczak}, {McCauliff}, {Morris}, \&
  {Twicken}}]{Smith:2012a}
{Smith}, J.~C., {Stumpe}, M.~C., {Van Cleve}, J.~E., {et~al.} 2012, \pasp, 124,
  1000

\bibitem[{{Welsh} {et~al.}(2007){Welsh}, {Wheatley}, {Seibert}, {Browne},
  {West}, {Siegmund}, {Barlow}, {Forster}, {Friedman}, {Martin}, {Morrissey},
  {Small}, {Wyder}, {Schiminovich}, {Neff}, \& {Rich}}]{Welsh:2007}
{Welsh}, B.~Y., {Wheatley}, J.~M., {Seibert}, M., {et~al.} 2007, \apjs, 173,
  673

\bibitem[{{West} {et~al.}(2008){West}, {Hawley}, {Bochanski}, {Covey}, {Reid},
  {Dhital}, {Hilton}, \& {Masuda}}]{West:2008}
{West}, A.~A., {Hawley}, S.~L., {Bochanski}, J.~J., {et~al.} 2008, \aj, 135,
  785

\bibitem[{{Wolfgang} \& {Laughlin}(2011)}]{Wolfgang:2011a}
{Wolfgang}, A., \& {Laughlin}, G. 2011, ArXiv e-prints

\bibitem[{{Zsom} {et~al.}(2013){Zsom}, {Seager}, {de Wit}, \&
  {Stamenkovic}}]{Zsom:2013}
{Zsom}, A., {Seager}, S., {de Wit}, J., \& {Stamenkovic}, V. 2013, ArXiv
  e-prints

\end{thebibliography}
\end{document}